\documentclass[iop]{emulateapj}
\usepackage{graphicx}
\usepackage{amssymb}
\usepackage{threeparttable}

\mathchardef\mhyphen="2D

\slugcomment{Accepted for publication in ApJ Letters 2013}

\shorttitle{Dust in M83}
\shortauthors{Liu et al.}

\begin{document}


\title{Extinction and Dust Geometry in M83 H \lowercase{\sc ii} Regions: A Hubble Space Telescope/WFC3 Study}


\author{
Guilin Liu\altaffilmark{1,2},
Daniela Calzetti\altaffilmark{1},
Sungryong Hong\altaffilmark{1,3}, 
Bradley Whitmore\altaffilmark{4}, 
Rupali Chandar\altaffilmark{5}, 
Robert W. O'Connell\altaffilmark{6},
William P. Blair\altaffilmark{2,7}, 
Seth H. Cohen\altaffilmark{8},
Jay A. Frogel\altaffilmark{9}
and
Hwihyun Kim\altaffilmark{8}
}

\affil{$^1$Astronomy Department, University of Massachusetts, Amherst, MA 01003, USA}
\affil{$^2$Center for Astrophysical Sciences, Johns Hopkins University, Baltimore, MD 21218, USA; liu@pha.jhu.edu}
\affil{$^3$National Optical Astronomy Observatory, Tucson, AZ 85719, USA}
\affil{$^4$Space Telescope Science Institute, Baltimore, MD 21218, USA}
\affil{$^5$Department of Physics and Astronomy, University of Toledo, Toledo, OH 43606, USA}
\affil{$^6$Astronomy Department, University of Virginia, P.O. Box 3818, Charlottesville, VA 22903, USA}
\affil{$^7$Visiting Astronomer, Las Campanas Observatory, La Serena, Chile}
\affil{$^8$School of Earth and Space Exploration, Arizona State University, Tempe, AZ 85287, USA}
\affil{$^9$Galaxies Unlimited, Lutherville, MD 21093, USA}



\begin{abstract}

We present HST/WFC3 narrow-band imaging of the starburst galaxy M83 targeting the hydrogen 
recombination lines (H$\beta$, H$\alpha$ and Pa$\beta$), which we use to investigate the 
dust extinction in the H {\sc ii} regions. We derive extinction maps with 6 parsec spatial 
resolution from two combinations of hydrogen lines (H$\alpha$/H$\beta$ and H$\alpha$/Pa$\beta$), 
and show that the longer wavelengths probe larger optical depths, with $A_V$ values larger 
by $\gtrsim$1 mag than those derived from the shorter wavelengths. 
This difference leads to a factor $\gtrsim$2 discrepancy in the extinction-corrected H$\alpha$ 
luminosity, a significant effect when studying extragalactic H {\sc ii} regions. 
By comparing these observations to a series of simple models, we conclude that a large 
diversity of absorber/emitter geometric configurations 
can account for the data, implying a more complex physical structure than the classical 
foreground ``dust screen'' assumption. 
However, most data points are bracketed by the foreground screen and a model where dust 
and emitters are uniformly mixed. 
When averaged over large ($\gtrsim$100--200 pc) scales, the extinction becomes consistent 
with a ``dust screen'', suggesting that other geometries tend to be restricted 
to more local scales. 
Moreover, the extinction in any region can be 
described by a combination of the foreground screen and the uniform mixture model 
with weights of 1/3 and 2/3 in the center ($\lesssim$2 kpc), respectively, and 2/3 and 
1/3 for the rest of the disk. This simple prescription significantly improves 
the accuracy of the dust extinction corrections and can be especially useful for 
pixel-based analyses of galaxies similar to M83.

\end{abstract}


\keywords{galaxies: individual (M83, NGC 5236) --- galaxies: ISM --- dust, extinction}



\section{Introduction}
\label{sec:intro}

Interstellar dust attenuates and reddens the light from stars, by both absorbing and 
scattering it, producing what is termed ``dust extinction''. When dealing with individual 
stars, one can usually assume that the dust (absorber) 
is located exclusively in front of the star (emitter) and is well separated from it,
thus simplifying the treatment of dust extinction. The analysis of the light-attenuating effects of 
dust quickly becomes extremely complicated when extended sources, such as stellar 
populations, are the emitters, and absorbers and emitters can distribute to produce complex geometries. 
This is generally the case of external galaxies, where individual stars are unresolved; in this case we term 
``dust attenuation'' the combined effects of extinction and geometry. The
assumption that the absorbers are located in front of the emitters and well separated from 
them is widely applied in astronomy even for external galaxies, but is far from accurate.
	
Probing the actual geometry of the dust distribution in an external galaxy is a challenging 
task that requires multiple-band measurements across a wide range of the electromagnetic 
spectrum with very high spatial resolution.  
The hydrogen recombination emission lines provide a convenient tool, commonly used for
deriving both the dust extinction and the dust geometry.  The intrinsic ratios of these 
lines are easily calculated theoretically, and show little variation for a wide range 
of physical and chemical conditions. For case B recombination, changes 
in the electron density of an H {\sc ii} region between  $10^2$ and  $10^4$ cm$^{-3}$ and 
in the temperature between $5\times10^3$ and  $10^4$ K, the H$\alpha$/H$\beta$ line ratio 
only changes between 3.04 and 2.85 ($\sim$7\%), and H$\alpha$/Pa$\beta$ between 
16.5 and 17.6 \citep[also $\sim$7\%,][]{Osterbrock06,Dopita03}. Case B recombination and the 
adopted density and temperature ranges are appropriate for most extragalactic nebulae. 
Observing two hydrogen lines yields a crude measurement of the foreground extinction 
(similar to what is done for individual stars), but at least three or more lines, ideally 
widely spaced in wavelength, are needed to constrain both extinction and geometry.

The starburst galaxy M83 has been imaged in multiple narrow  bands targeting hydrogen recombination 
lines (H$\beta$ and H$\alpha$ in the optical, and Pa$\beta$ in the near-IR), as part of the Early Release 
Science (ERS) observations made by the HST/WFC3 Scientific Oversight Committee (SOC), which enables 
an unprecedented detailed study of the dust extinction and geometry in the nebular gas of a nearby galaxy.	
	
M83 (NGC 5236, a.k.a. the ``Southern Pinwheel Galaxy''), is classified as an SAB(s)c \citep[][]{RC3}, 
is one of the closest grand design-spirals (4.56 Mpc), is virtually face-on, and has a starburst nucleus, 
a marked bar, and prominent spiral arms. These properties make it an ideal target for detailed investigations 
of dust extinction in extragalactic H {\sc ii} regions.


\section{Data Processing}

M83 was mapped with HST/WFC3 in 2009 August, with a series of narrow and broad 
band filters in the UVIS and IR channels, as part of the ERS program (GO-11360, 
PI: Robert O'Connell; see \citealt{Chandar10} for further details of the program 
and the full list of filters used to observe M83). 
The narrow-band filters included those in the H$\beta$ ($\lambda$4861\AA, F487N), 
H$\alpha$+[N {\sc ii}] ($\lambda$6563\AA+$\lambda\lambda$6548, 6584\AA, F657N) and 
Pa$\beta$ ($\lambda$12818\AA, F128N) emission lines. 
The observations cover two pointings: the southern pointing targets 
the nuclear region of M83 and about half of its northeastern arm, and the northern pointing targets an
adjacent field of the same area where the southwestern arm winds over.
The field of view (FoV) for each pointing subtends 3.6$\times$3.6 kpc$^2$ for H$\alpha$ and 
H$\beta$, and 3.1$\times$2.7 kpc$^2$ for Pa$\beta$. 

The raw data were first processed with the {\sl MultiDrizzle} software 
\citep{Fruchter09} to accomplish both basic and high-level processing, including flat-fielding, 
cosmic-ray cleaning, combination, mosaicing of the two pointings, and registration 
of all mosaics onto a common grid \citep{Mutchler10}. 

We proceed at matching the point spread functions (PSFs) of the UVIS and IR channels. Direct measurements 
of point sources in the images yield PSFs with 0\farcs07, 0\farcs07 and 0\farcs22 FWHM for 
H$\beta$, H$\alpha$, and Pa$\beta$, respectively. We thus degrade the UVIS images 
(0\farcs07 resolution) to match the resolution of IR data (0\farcs22) by convolving 
the former to a Gaussian kernel. Finally, we resample the images from both channels 
(original pixel sizes are 0\farcs04 for UVIS and 0\farcs13 for IR) to a 7$\times$7 grid of the original UVIS 
data, obtaining a final pixel size of 0\farcs28 (6.1 pc).

To produce pure hydrogen emission line images, we subtract a rescaled F110W image from the 
F128N image to remove the stellar continuum. The stellar continuum images for H$\beta$ and H$\alpha$ 
lines are determined by linearly interpolating between the
F438W and F555W and between the F555W and F814W images, respectively. For each individual 
pointing in each line, we subtract the stellar continuum by matching the fluxes 
of a sample of stars, and then measure and remove the global background.
Filter throughput corrections for the shifted lines at the  513 km s$^{-1}$ heliocentric 
recession velocity of M83 \citep{Koribalski04} amount to only 1--3\%. Each emission line 
image is also corrected for foreground Galactic extinction using the color excess value 
$E(B-V)$=0.066 mag, as reported in 
NED\footnote{http://ned.ipac.caltech.edu. The NASA/IPAC Extragalactic Database (NED) is operated by the Jet Propulsion Laboratory, California Institute of Technology, under contract with the National Aeronautics and Space Administration.}. 

\begin{figure*}
\centering
\includegraphics[scale=1.05,trim=1cm 0mm 2cm 0mm]{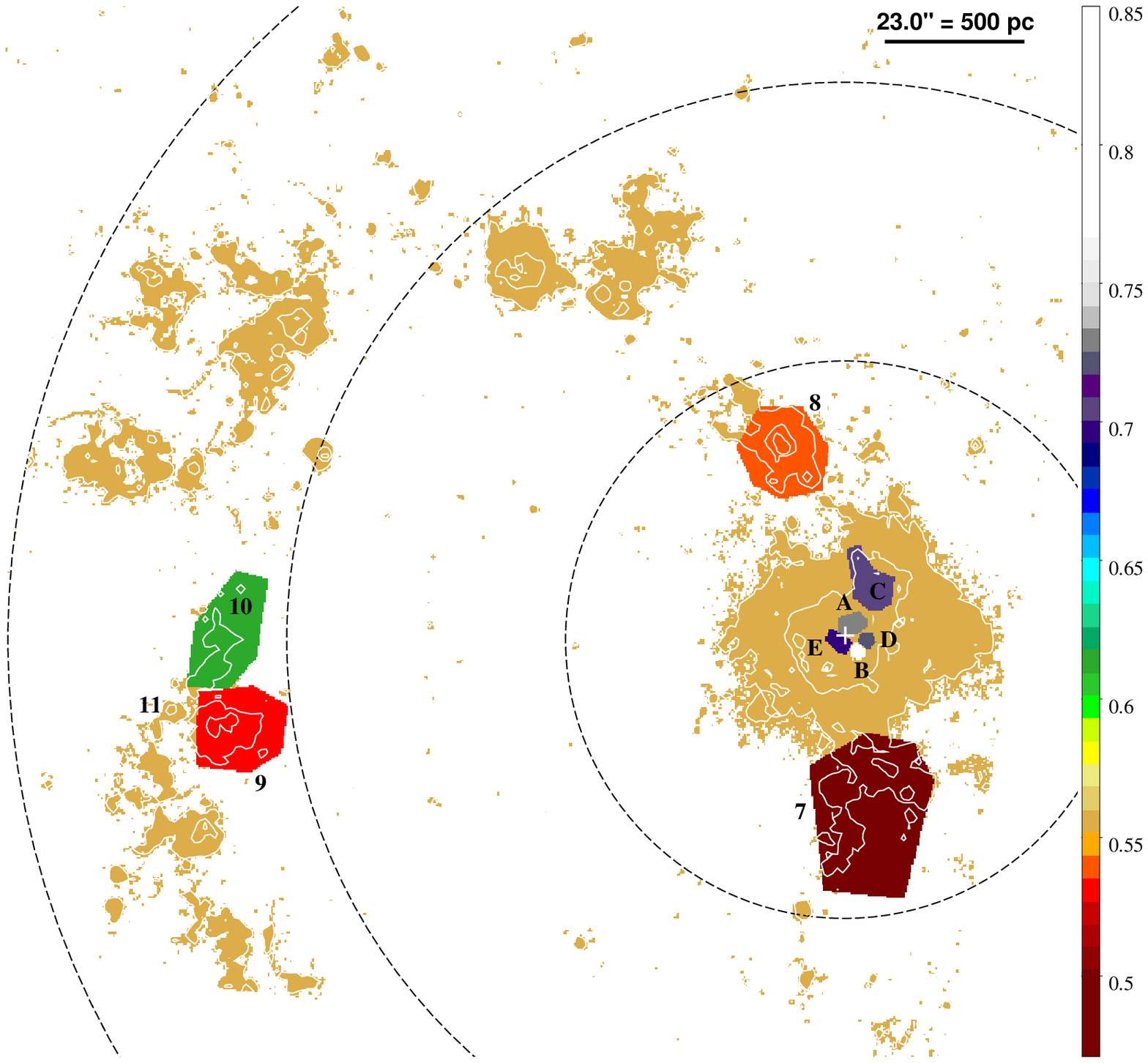}
\caption{The [N {\sc ii}]/H$\alpha$ ratio map for M83 adopted in this work, 
overlaid on the H$\alpha$+[N {\sc ii}] contour map. The letters and numbers 
denote the regions studied by \citet{Bresolin02}. Only the southern pointing 
is shown because none of the H {\sc ii} regions in the northern pointing 
has been observed by those authors. The color bar to the right indicates the 
ratio of the total intensity of the nitrogen doublet 
($\lambda\lambda$6548, 6584\AA) to the H$\alpha$ line intensity. The kinematic
nucleus is marked with a white plus sign following \citet{Knapen10}, 
and the black dashed circles depict the galactic radii of 1, 2 and 3 kpc 
that are used for the calculations in Table \ref{tab1} and Figure \ref{fig:Av_hist}.}
\label{fig:n2_corr}
\end{figure*}

The F657N filter allows [N {\sc ii}]$\lambda\lambda$6548, 6584\AA\ doublet, which needs to 
be removed to obtain the H$\alpha$ emission. To account for spatial variations in the 
[N {\sc ii}]/H$\alpha$ ratio, we use the measurements in \citet{Bresolin02}, who have 
obtained spectroscopic data for 11 H {\sc ii} regions in the disk and
5 regions in the core of this galaxy. Our FoV covers 5 of the disk regions
and all the nuclear regions. We assign the spectroscopic  [N {\sc ii}]/H$\alpha$ ratio 
determinations to areas surrounding each H {\sc ii} region from the \citet{Bresolin02} paper.  
Figure \ref{fig:n2_corr} shows the distribution of the adopted ratios, color coded in terms of the total 
([N {\sc ii}]$\lambda$6548\AA+[N {\sc ii}]$\lambda$6584\AA)/H$\alpha$ ratio. 
For the areas outside the immediate surroundings of each H {\sc ii} region and for the entire 
northern pointing \citep[which does not contain any H {\sc ii} region studied by][]{Bresolin02}, 
we assume the value of Region 11 in their paper, the one closest to the average of the four regions 
located on the northeastern spiral arm. 

\begin{figure*}
\centering
\includegraphics[scale=1.1,trim=9cm 0mm 11cm 0mm]{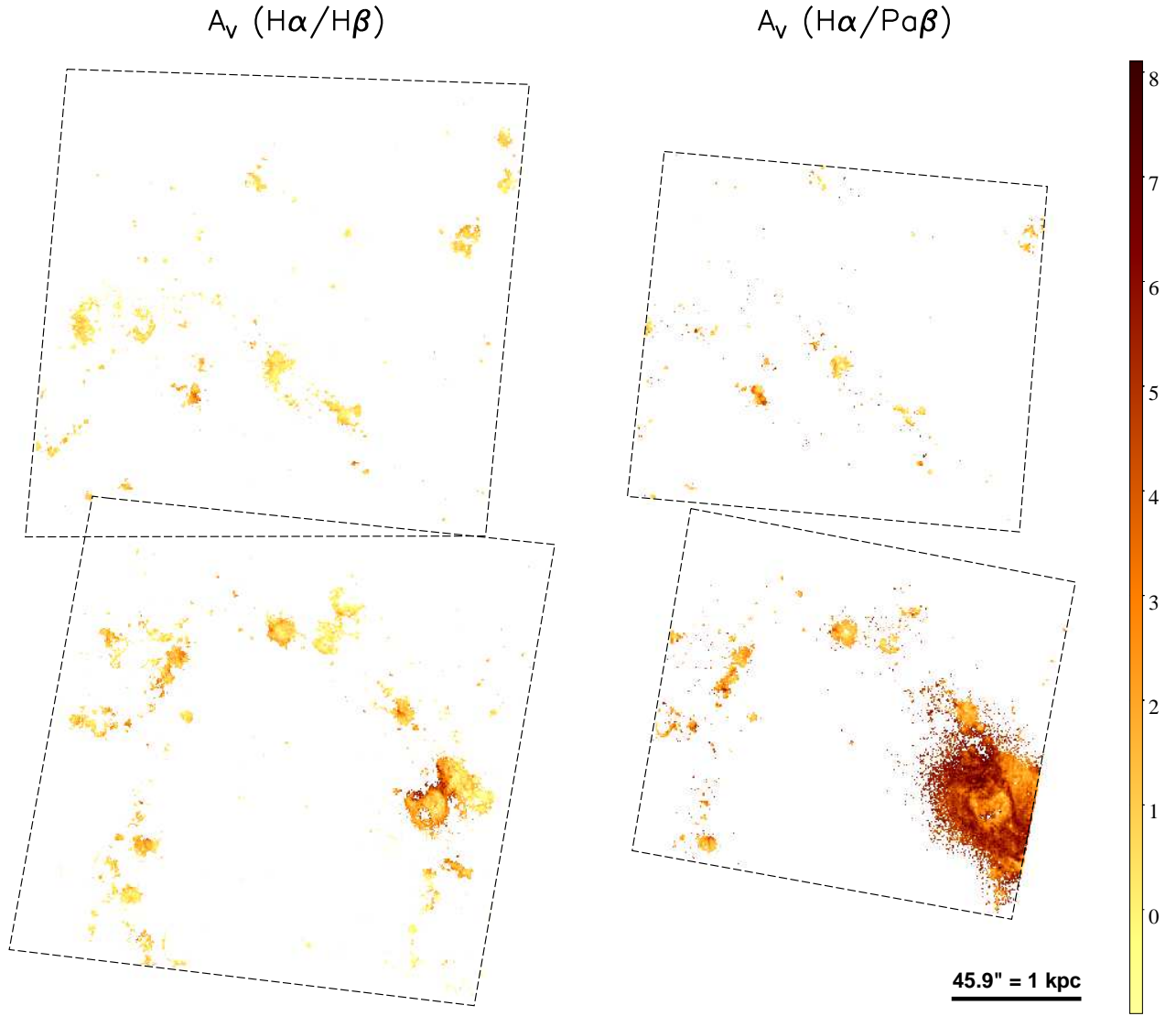}
\caption{The visual extinction ($A_V$) maps in M83. These two $A_V$ maps are derived from the 
H$\alpha$/H$\beta$ (left) and H$\alpha$/Pa$\beta$ (right) line ratios. The southern and northern
pointings have been mosaiced, and the colorbar shows the value of $A_V$ in magnitudes. The
margins of the FoVs are depicted by the black dash lines for both the northern and the southern 
pointings.}
\label{fig:Av_maps}
\end{figure*}

\begin{center}
\begin{deluxetable*}{c c c c c c}
\setlength{\tabcolsep}{0.02in} 
\tablecaption{Derived dust properties in M83. \label{tab1}}
\tablehead{
\colhead{Region} 
& \colhead{$A_V{\rm (\frac{H\alpha_{tot}}{H\beta_{tot}})}$} 
& \colhead{$A_V{\rm (\frac{H\alpha_{tot}}{Pa\beta_{tot}})}$} 
& \colhead{$\langle A_V{\rm (\frac{H\alpha}{H\beta})}\rangle$} 
& \colhead{$\langle A_V{\rm (\frac{H\alpha}{Pa\beta})}\rangle$} 
& \colhead{$\log\,\langle N_{\rm d}\rangle$} \\
\colhead{(1)} & \colhead{(2)} & \colhead{(3)} 
& \colhead{(4)} & \colhead{(5)} 
& \colhead{(6)}
}

\startdata

all       &  1.48  &  2.65  & 0.61 &  4.04  &  10.07  \\
0--1 kpc  &  1.75  &  3.40  & 1.15 &  4.65  &  10.13 \\
1--2 kpc  &  0.82  &  2.65  & 0.23 &  1.58  &   9.67 \\
2--3 kpc  &  1.37  &  1.46  & 0.72 &  1.88  &   9.74 \\
3--4 kpc  &  1.33  &  0.68  & 0.50 &  1.14  &   9.53 \\
4--5 kpc  &  1.42  &  0.02  & 0.31 &  1.09  &   9.50 

\enddata

\tablecomments{
(1) Region for calculation.
(2--3) Visual extinction derived from the ratio of the total line emission from the considered region (mag).
(4--5) Median value of the $A_V$ maps (i.e., Figure \ref{fig:Av_maps}) within this region (mag).
(6) Dust Column density converted from Column 5 (cm$^{-2}$).
Note that the method for columns 4--6 is for high resolution studies, while the calculation for
columns 2--3 is relevant for analyses where only coarse resolution images (e.g. those of distant galaxies) 
are available. 
}
\end{deluxetable*}
\end{center}

\section{Results}

For our galaxy with super-solar metallicity \citep[$12+\log {\rm [O/H]}=8.94$,][]{Bresolin02,Hong13},
we adopt an electron temperature $T_{e}$=7500 K and a number density $n_{e}$=10$^3$ cm$^{-1}$ 
as the fiducial environment in the whole galaxy, implying intrinsic ratios 
$I_{\rm H\alpha}/I_{\rm H\beta}$=2.92 and $I_{\rm H\alpha}/I_{\rm Pa\beta}$=17.1 \citep{Dopita03} 
(note that our results are not sensitive to the specific choice of the electron density 
or temperature, see Section \ref{sec:intro}).
We adopt the extinction curve presented by \citet{Cardelli89}, with $k$(H$\alpha$)=2.535, 
$k$(H$\beta$)=3.609 and $k$(Pa$\beta$)=0.840, 
in the expression $I_{\rm obs}/I_{\rm intr}=10^{-0.4k(\lambda)E(B-V)}$.
Two extinction maps are then created, one derived from the ratio of H$\alpha$ to H$\beta$ and 
the other from that of H$\alpha$ to Pa$\beta$, as shown in Figure~\ref{fig:Av_maps}, using only 
pixels above the 5$\sigma$ detection level in all three maps. Our main limitation is the 
depth of the H$\beta$ line, which is the most dust attenuated one among the three. 

\subsection{Extinction and dust content}

\begin{figure*}
\centering
\includegraphics[scale=0.65, trim=0mm 0mm 0mm 0mm]{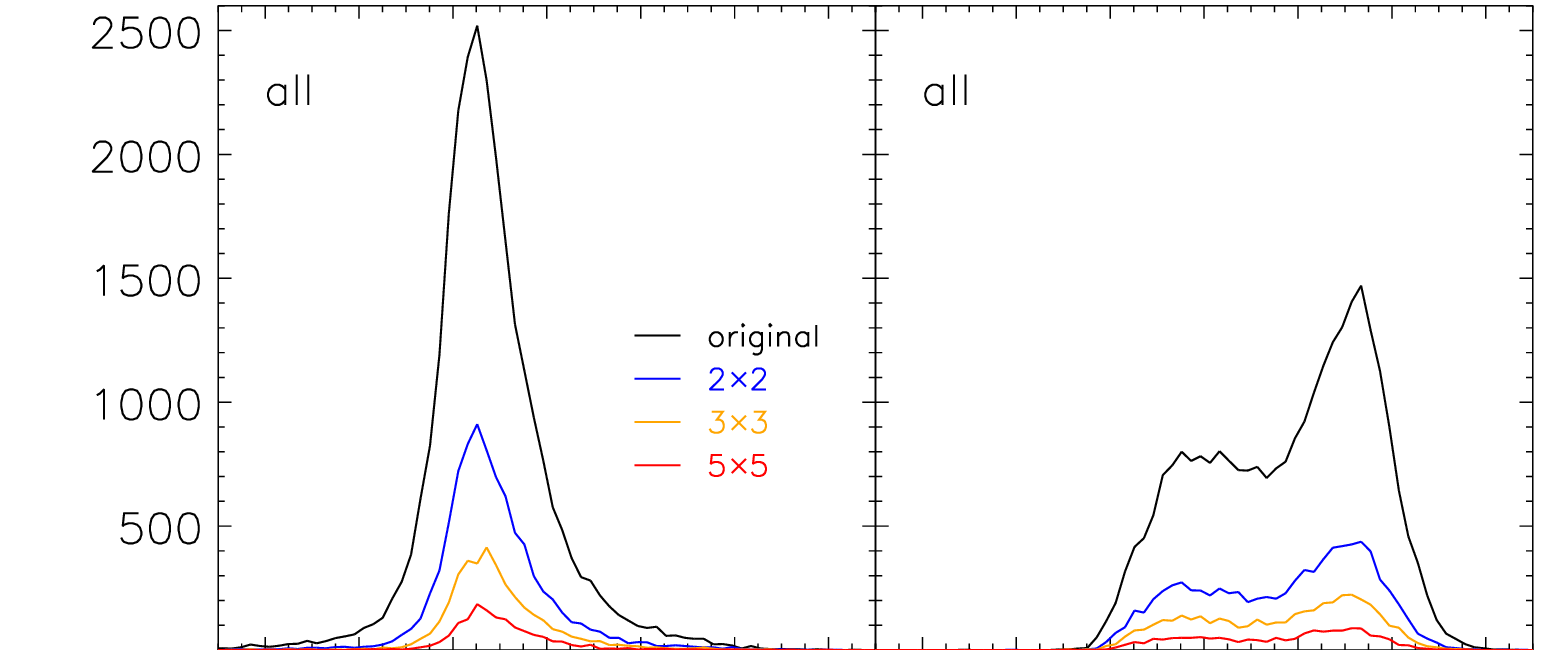}\\
\vspace{1mm}
\includegraphics[scale=0.65, trim=0mm 0mm 0mm 0mm]{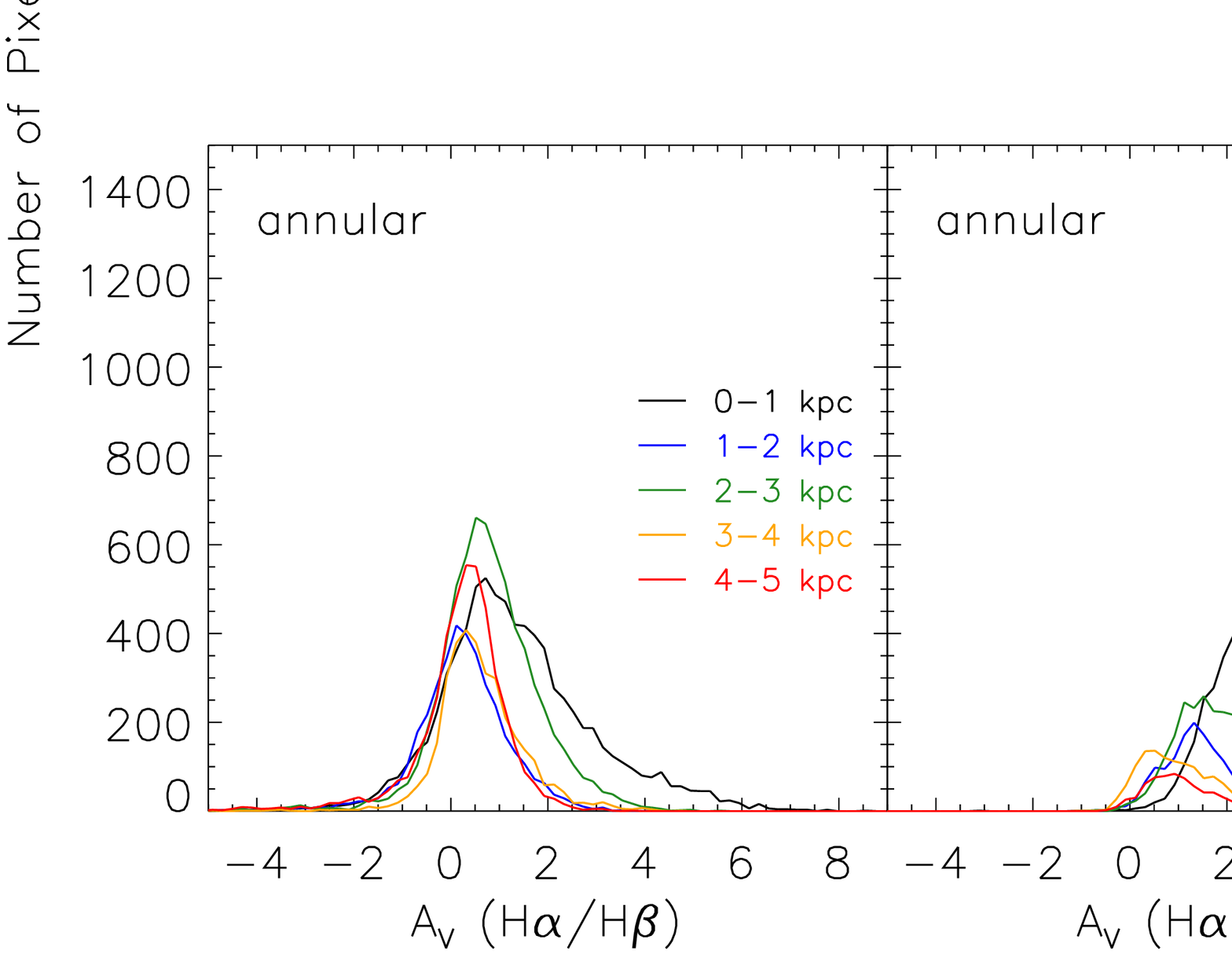}
\caption{Histograms of the visual extinction $A_V$ in the whole FoV (top) and a series of
annuli (bottom), derived from the H$\alpha$/H$\beta$ and H$\alpha$/Pa$\beta$ line ratios. 
The difference in the number of the involved pixels is a result of their respective 5-$\sigma$ cutoffs and FoVs. The histograms as a result of rebinning the maps onto a coarser grid are over-plotted in the top panels.}
\label{fig:Av_hist}
\end{figure*}

Figure~\ref{fig:Av_hist} shows the histograms of $A_V$ derived from the two line ratios.
A significant number of pixels in the H$\alpha$/H$\beta$ ratio map have negative values of $A_V$, 
due to the large error bars (typically $\sim$1 mag). In general, $A_V$ from H$\alpha$/H$\beta$ 
shows a narrow distribution peaked at $\sim$0.5 mag, while $A_V$ from H$\alpha$/Pa$\beta$ 
shows a double peak at $\sim$2 and $\sim$5.5 mag. Rebinning our maps with a coarser
grid hardly changes the shapes and the peak locations of the histograms (cf. the color
lines in the figure). We also show these histograms in a series of galactic annuli, which
clearly demonstrate that $A_V$ (H$\alpha$/Pa$\beta$) peaks at a larger value than 
$A_V$ (H$\alpha$/H$\beta$) in every annulus (also see Table \ref{tab1}), and the second peak of $A_V$ (H$\alpha$/Pa$\beta$)
mentioned above is largely contributed by the nucleus, where many pixels with large and 
patchy dust attenuation are present \citep[e.g.,][]{Thatte00}. 

Although the size of dust grains spans a wide range, the extinction in $V$-band ($\lambda$=5500\AA) 
is dominated by the grains with a radius $a\sim\lambda/2\pi\sim0.1$ $\mu$m, thus we can derive 
the dust column density using $A_V=1.086\,\tau_V=1.086\,\pi a^2 N_{\rm d}$, and
further calculate the dust mass by assuming the solid density of the grain material to be  
$\rho_{\rm d}=3$ g cm$^{-3}$, a compromise between graphite and crystalline olivine \citep{Draine11}.
Using $A_V$ (H$\alpha$/Pa$\beta$) which is likely closer to the actual extinction than $A_V$ (H$\alpha$/H$\beta$), 
we find the dust mass in the whole FoV to be $7\times10^5$ $M_{\odot}$.
The entire galaxy has $\sim$24 times of this amount ($\sim2\times10^7$ $M_{\odot}$),
as our FoV covers $\sim$8\% of the disk ($12.9\arcmin\times11.5\arcmin$, NED), and a factor 
of 2 is to account for the dust in the excluded S/N$<$5 regions. 
This estimation, although an upper limit (these faint regions likely do not contribute that 
much, and the galaxy center, the most dusty region, is covered by our FoV),
is consistent with previous studies using infrared and (sub)millimeter data 
(\citealt{Galametz11} find $8.5\times10^6~M_{\odot}$; \citealt{Foyle12} find $4\times10^7~M_{\odot}$),
and is similar to the dust content of other metal-rich spiral galaxies \citep[e.g.,][]{Liu10,Draine07}.

\subsection{Dust geometry}

\begin{figure*}
\centering
\hspace{5mm}
\includegraphics[scale=0.5]{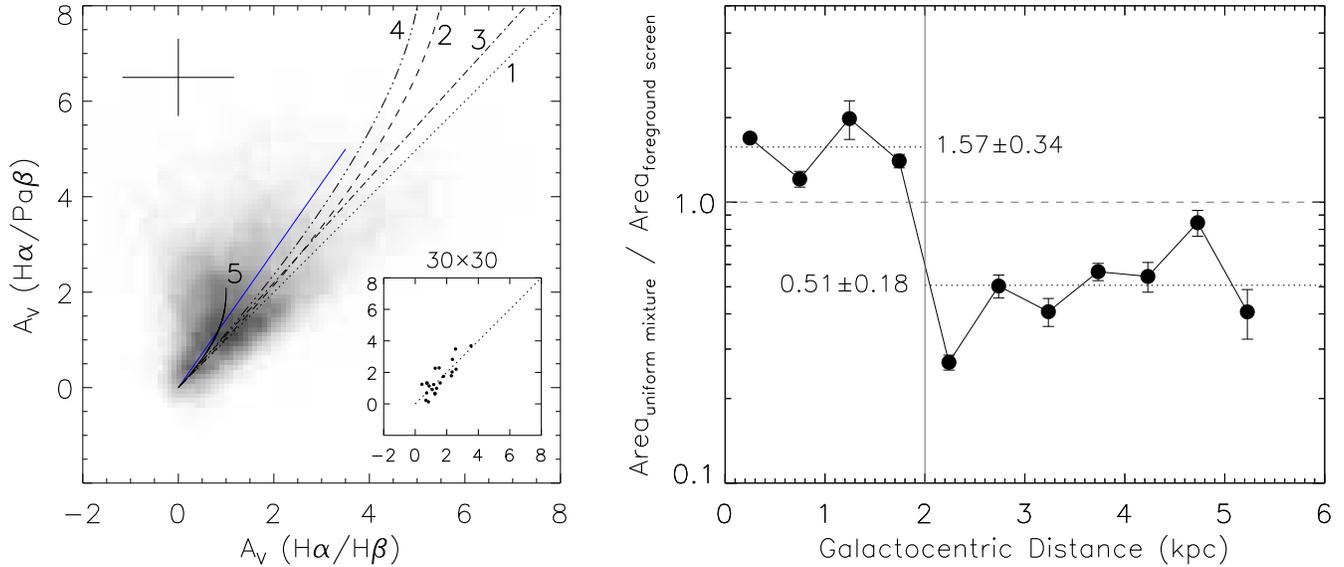}
\vspace{1.5cm}
\caption{{\bf \itshape Left.} The relation between $A_V$ derived in two ways, compared to the 5 simple models
discussed in the text, numbered in the same fashion (1 -- foreground screen; 2 -- clumpy dust screen; 
3 -- uniform scattering slab; 4 -- clumpy scattering slab; 5 -- uniform mixture). 
Only pixels with S/N$\geqslant$5 detection in all three lines are used, and the typical 1-$\sigma$ 
uncertainty is shown at the corner.
The blue line $A_V{\rm(H\alpha/Pa\beta)}=1.428\;A_V{\rm(H\alpha/H\beta)}$ bisects the data points and
approximately separates the two branches which roughly follow either model 5 or the 
degenerate trend of the other four models. When averaged over $\sim$180 pc (i.e., a
$30\times30$ grid), all data points become consistent with model 1.
{\bf \itshape Right.} The ratio of the area occupied by the ``uniform mixture branch'' to that by 
the ``foreground screen branch'' as a function of the galactocentric distance. 
The central region within $\sim$2 kpc is dominated by ``uniform mixture'' pixels (61\%) and thus favors model 5, 
while the rest of the disk is dominated by ``foreground screen'' pixels (66\%) and favors model 1 better.}
\label{fig:models}
\end{figure*}

The availability of three hydrogen emission lines also enables a crude analysis
of the dust geometry in the H {\sc ii} regions, by comparing the observationally derived extinction
to models of the emitters/absorbers geometry. The simplest geometry, 
the foreground screen, in which the dust grains form a homogeneous, non-scattering screen 
foreground to the emitting light source, is such that any pair of hydrogen lines will yield the same $A_V$ value. 
This is clearly not the case in M83. When the $A_V$ values derived from H$\alpha$/H$\beta$ and from H$\alpha$/Pa$\beta$ are compared to each other, as shown in Figure~\ref{fig:models}, they do not follow a relation with a slope of unity, 
as expected in the case of the foreground screen. Instead, the data points are broadly distributed
in the upper half of the diagram, delimited by the one-to-one line and the vertical
$A_V=0$ line. 

In order to describe the behavior observed in Figure \ref{fig:models}, we follow \citet{Calzetti94} 
and \citet{Natta84} and calculate the predictions of five simple models with plane-parallel 
absorber/emitter configurations \citep[for a schematic representation of the models, see][]{Calzetti94}:

\begin{enumerate}
\item The foreground screen model, as described above (no clumpiness and 
scattering is present, as the screen is uniform and physically distant from the light emitter).
\item The clumpy dust screen model. Like \citet{Natta84} and \citet{Calzetti94}, we assume all the 
clumps to have the same optical depth and to be Poisson distributed, with an average of $N$ clumps
along the line of sight.
\item The uniform scattering slab model. In this case the dust is located close to the 
emitter and the scattering by the dust grains into the line of sight has an important positive
contribution. We follow the calculation procedure of \citet{Calzetti94}, but with the
physical parameters of the dust from \citet{Witt00}.
\item The clumpy scattering slab model. The properties of this model is a combination of
models 2 and 3. Same dust parameters are used as in model 3. The same value of $N$ as model
2 has been adopted.
\item The uniform mixture model. The dust and emitters are homogeneously mixed together, and the 
scattering into the line of sight is taken into account. 
\end{enumerate}

The relations between the visual extinction derived from these hydrogen lines predicted by 
the above models are shown in Figure \ref{fig:models}. 
The average number of clumps in models 2 and 4 are set to be $N=6$ in the figure. Model 5 has the 
interesting property that the tip of the line is the limit when the optical depth becomes 
infinitely large, in which case neither of the three emission lines can explore deep into the 
H {\sc ii} regions. None of the five models can, alone, account for the distribution 
of all data points, even when taking the fairly large error bars into account. The spread in 
extinction values is significant and spans the range covered by all 5 models. A significant 
number (30.4\%) of regions with $A_V~\rm (H\alpha/H\beta) < 2$ mag have extinction values 
$A_V>2$ mag when the latter is derived from H$\alpha$/Pa$\beta$. For these regions, the 
traditional use of the H$\alpha$/H$\beta$ line ratio to derive extinction values produces a 
significant underestimate of the intrinsic values, by factors $>$2 (up to $\sim$10).
This is in line with the conclusions of \citet{Israel80} and \citet{Skillman88} that deriving dust extinction 
in H {\sc ii} regions at longer wavelengths results in larger values than at shorter wavelengths.

The extinction values are almost evenly distributed between the two extreme lines 
marked by models 1 and 5 (and even beyond those models), suggesting a large variation in the 
dust geometry among different H {\sc ii} regions. At the same time, these models clearly 
encompass the majority of the data, when the data uncertainties are taken into account. 
The envelope represented by model 1 and model 5 suggests 
that for most H {\sc ii} regions/nebulae, the uniform mixture and the foreground screen models 
bracket the range of absorber/emitter geometric configurations.

Although clearly an over-simplification of the complex dust geometry, 
model 1 represents 46\% of the area of H {\sc ii} regions well (within 1/2 magnitude of extinction). 
The most deviant points are located in the nuclear regions of this galaxy.

Moreover, rebinning our maps onto a very coarse ($\gtrsim 20\times20$) 
grid significantly suppresses the necessity to introduce model 5. For instance, at a sampling scale 
of $\sim$180 pc ($30\times30$) all data points agree with a dust screen configuration 
(Figure \ref{fig:models}). This is consistent with \citet{Kennicutt09}: when averaged over 
large sub-galactic regions, the mean extinction becomes consistent with the foreground dust 
screen model, suggesting that other geometries tend to be more localized.

\section{Potential applications}

In spite of the significant uncertainty of the derived $A_V$, the data points in Figure \ref{fig:models} 
appear to form two branches: one of which approximately follows the locus of model 5 (the 
``uniform mixture branch''), while the other favors models 1--4 whose degeneracy between each 
other is difficult to break and therefore can be represented by model 1 (the ``foreground screen branch''). 
Projecting these two branches to the M83 images will provide an empirical prescription for extinction 
correction in spatially resolved galaxies.

These two branches can be reasonably divided by the straight line that bisects all the data points 
$A_V{\rm(H\alpha/Pa\beta)}=1.428\;A_V{\rm(H\alpha/H\beta)}$ (the blue line in Figure \ref{fig:models}). 
To investigate the spatial distribution of the data points separated by this line, we calculate the 
ratio of the area occupied by the points from each branch within a series of annuli with a fixed width 
of 0.5 kpc starting from the galaxy center. Interestingly, we find this ratio to persist at roughly 
a constant on either side of a radius of $\sim$2 kpc, but abruptly switch their roles at this radius
(Figure \ref{fig:models}). Specifically, in terms of the mean values and standard deviations, we find
\begin{equation}
 \left<\frac{{\rm Area_{uniform~mixture}}}{{\rm Area_{foreground~screen}}}\right> = \left\{ 
  \begin{array}{l l} 
     1.57\pm0.34 ~~~~~~~{\rm (\lesssim2~kpc)}; \\
     0.51\pm0.18 ~~~~~~~{\rm (>2~kpc)}.  \\
  \end{array} \right. 
\end{equation}

Hence, we conclude that in the central region of M83 within a galactocentric distance of $\sim$2 kpc, 
61\% of the nebular-emitting area follows the dust extinction properties of model 5 (uniform mixture). 
Beyond this radius, 66\% of the nebular-emitting area follows the foreground screen model. 
This result is particularly useful for pixel-based statistical analyses of
spiral galaxies with properties similar to M83, for which we suggest that models 5 (1) be used with a
weight of $\sim$1/3 (2/3) for the center and $\sim$2/3 (1/3) for the outer disk. Even a more crude
prescription --- emitters and absorbers are assumed to be uniformly mixed in the center 
($\lesssim 2$ kpc) and the foreground dust screen is employed for the outer disk --- will significantly
improve the accuracy of dust extinction correction.

\section{Summary}

As part of the Early Release Science observations made by the HST/WFC3 Scientific 
Oversight Committee, the starburst galaxy M83 has been imaged in multiple narrow 
bands targeting three hydrogen recombination lines (H$\beta$, H$\alpha$ and Pa$\beta$).
These data enable us to scrutinize the extinction values and geometry in the H {\sc ii} regions 
by deriving the $A_V$ map using two combinations of hydrogen lines: H$\alpha$/H$\beta$ 
and H$\alpha$/Pa$\beta$. The pixel-by-pixel comparison between the two extinction maps  
shows that  larger optical depths are probed by the longer wavelength line emission,  
yielding $A_V$ values that are larger by $\gtrsim$1 mag than those derived from shorter-wavelength 
hydrogen lines. This produces a factor $\gtrsim$2 discrepancy in the intrinsic H$\alpha$ luminosity 
when using the different line ratios for the extinction correction. By comparing these observations 
to a series of simple models, we conclude that the data require a large diversity of absorber/emitter 
geometric configurations, but when averaged over large ($\gtrsim$100--200 pc) sub-galactic regions, 
the mean extinction becomes consistent with the foreground dust screen model, suggesting that other 
geometries tend to be restricted to more local scales. 
Moreover, we can provide a simple prescription for improving the extinction 
corrections in spatially-resolved analyses, by expressing the correction in terms of two extreme 
geometrical dust configurations: (1) foreground non-scattering dust screen; and (2) uniform mixture of 
emitters and absorbers. These two configurations can be combined in the following proportions: 
$\sim$2/3 (1/3) for uniform mixture (foreground screen) in the central area ($\lesssim 2$ kpc) 
and $\sim$1/3 (2/3) in the outer disk.

\acknowledgments

This paper is based on Early Release Science observations made by the WFC3 Science Oversight Committee. 
Support for program GO--11360 was provided by NASA through a grant from Space Telescope Science 
Institute, which is operated by the Association of Universities for Research in Astronomy, Inc., 
under NASA contract NAS 5-26555. 
We are grateful to the Director of STScI for awarding Director’s Discretionary Time for this program.



{\it Facilities:} \facility{HST (WFC3)}.


\begin{thebibliography}{}
\bibitem[Bresolin \& Kennicutt(2002)]{Bresolin02} Bresolin, F., \& Kennicutt, R.~C., Jr.\ 2002, \apj, 572, 838 
\bibitem[Calzetti et al.(1994)]{Calzetti94} Calzetti, D., Kinney, A.~L., \& Storchi-Bergmann, 
	T.\ 1994, \apj, 429, 582 
\bibitem[Cardelli et al.(1989)]{Cardelli89} Cardelli, J.~A., Clayton, G.~C., \& Mathis, 
	J.~S.\ 1989, \apj, 345, 245 
\bibitem[Chandar et al.(2010)]{Chandar10} Chandar, R., et al.\ 2010, \apj, 719, 966 
\bibitem[de Vaucouleurs et al.(1991)]{RC3} de Vaucouleurs, G., de Vaucouleurs, A., Corwin, H. G.,
        Buta, R. J., Paturel, G., Fouque, P., 1991, Third Reference Catalogue of Bright Galaxies, 
        Springer-Verlag, Berlin, Heidelberg, New York
\bibitem[Dopita \& Sutherland(2003)]{Dopita03} Dopita, M.~A., \& Sutherland, R.~S.\ 2003, Astrophysics 
	of the diffuse universe, Berlin, New York: Springer, 2003 
\bibitem[Draine et al.(2007)]{Draine07} Draine, B.~T., Dale, D.~A., Bendo, G., et al.\ 2007, \apj, 663, 866 
\bibitem[Draine(2011)]{Draine11} Draine, B.~T.\ 2011, Physics of the Interstellar and Intergalactic Medium,
	Princeton, New Jersey: Princeton Univ. Press, 2011 
\bibitem[Fruchter et al.(2009)]{Fruchter09} Fruchter, A., et al. 2009, The MultiDrizzle Handbook, 
	Version 3.0 (Baltimore, MD: STScI)
\bibitem[Foyle et al.(2012)]{Foyle12} Foyle, K., Wilson, C.~D., Mentuch, E., et al.\ 2012, \mnras, 421, 2917 
\bibitem[Galametz et al.(2011)]{Galametz11} Galametz, M., Madden, S.~C., Galliano, F., et al.\ 2011, \aap, 532, A56 
\bibitem[Hong et al.(2013)]{Hong13} Hong, S., Calzetti, D., Gallagher, J.~S., III, et al.\ 2013,
	\apj, in press (arXiv:1309.0520)
\bibitem[Israel \& Kennicutt(1980)]{Israel80} Israel, F.~P., \& Kennicutt, R.~C.\ 1980, \aplett, 21, 1 
\bibitem[Kennicutt et al.(2009)]{Kennicutt09} Kennicutt, R.~C., Jr., Hao, C.-N., Calzetti, D., et al.\ 
	2009, \apj, 703, 1672 
\bibitem[Knapen et al.(2010)]{Knapen10} Knapen, J.~H., Sharp, R.~G., Ryder, S.~D., et al.\ 
	2010, \mnras, 408, 797 
\bibitem[Koribalski et al.(2004)]{Koribalski04} Koribalski, B. S., Staveley-Smith, L., Kilborn, V. A., et al. 
	2004, \aj, 128, 16
\bibitem[Liu et al.(2010)]{Liu10} Liu, G., Calzetti, D., Yun, M.~S., et al.\ 2010, \aj, 139, 1190 
\bibitem[Mutchler(2010)]{Mutchler10} Mutchler, M. 2010, in Space Telescope Science Institute Calibration
	Workshop—Hubble after SM4, Preparing JWST, ed. S. Deustua \& C. Oliveira 
	(Baltimore, MD: Space Telescope Science Institute), 69
(http://www.stsci.edu/institute/conference/cal10/proceedings)
\bibitem[Natta \& Panagia(1984)]{Natta84} Natta, A., \& Panagia, N.\ 1984, \apj, 287, 228
\bibitem[Osterbrock \& Ferland(2006)]{Osterbrock06} Osterbrock, D.E. \& Ferland, G.J. 2006, Astrophysics \\
	of Gaseous Nebulae and Active Galactic Nuclei, 2nd. ed. (Sausalito, CA: Univ. Science Books)
\bibitem[Skillman \& Israel(1988)]{Skillman88} Skillman, E.~D., \& Israel, F. P. 1988, A\&A, 203, 226
\bibitem[Thatte, Tezca, \& Genzel(2000)]{Thatte00} Thatte, N., Tecza, M., \& Genzel, R. 2000, A\&A, 364, L47
\bibitem[Witt \& Gordon(2000)]{Witt00} Witt, A.~N., \& Gordon, K.~D.\ 2000, \apj, 528, 799 

\end{thebibliography}
\end{document}